\documentclass[12pt]{amsart}
\usepackage[english]{babel}
\usepackage{egothic}
\usepackage[T1]{fontenc}

\usepackage{amsmath}
\usepackage{amsthm}
\usepackage{amssymb}
\usepackage{amsfonts} 
\usepackage{mathrsfs} 
\usepackage{pb-diagram} 
\usepackage[all]{xy}
\usepackage{mathrsfs}
\usepackage{enumerate}
\usepackage{physics}
\usepackage[notcite, final, notref]{showkeys}
\usepackage{dsfont}
\usepackage[dvips]{color}
\newtheorem{theorem}{Theorem}[section]

\newtheorem{lemma}[theorem]{Lemma}
\newtheorem{proposition}[theorem]{Proposition}

\newtheorem{definition}[theorem]{Definition}

\theoremstyle{definition}
\newtheorem{remark}[theorem]{Remark}

\newtheorem{example}[theorem]{Example}

\newcommand{\BC}{{\mathbb B}{\mathbb C}}
\newcommand{\D}{{\mathbb D}}

\renewcommand{\i}{{\bf i}}
\renewcommand{\j}{{\bf j}}
\renewcommand{\k}{{\bf k}}
\newcommand{\C}{{\mathbb C}}
\newcommand{\R}{{\mathbb R}}
\newcommand{\e}{{\bf e}}

\renewcommand{\Re}{\mathrm{Re}}

\usepackage{xcolor}

\title[]{}

\author[D. Alpay]{Daniel Alpay}
\address{(DA) Schmid College of Science and Technology \\
Chapman University\\
One University Drive
Orange, California 92866\\
USA}
\email{alpay@chapman.edu}

\author[A. DeMartino]{Antonino De Martino}

\author[K. Diki]{Kamal Diki}
\address{(KD) Schmid College of Science and Technology \\
Chapman University\\
One University Drive
Orange, California 92866\\
USA}
\email{diki@chapman.edu}

\author[M. Vajiac]{Mihaela Vajiac}
\address{(MV) Schmid College of Science and Technology \\
Chapman University\\
One University Drive
Orange, California 92866\\
USA}
\email{mbvajiac@chapman.edu}

\title[Bicomplex Tensor Product and Choi Theorem]{The Bicomplex Tensor Product, a Bicomplex Choi Theorem and Applications}

\begin{document}
\maketitle
\begin{abstract}
In this paper we extend the concept of tensor product to the bicomplex case and use it to prove the bicomplex counterpart of the classical Choi theorem in the theory of complex matrices and operators. The concept of hyperbolic tensor product is also discussed, and we link these results to the theory of quantum channels in the bicomplex and hyperbolic case, as well as applications to bicomplex digital signal processing.
\end{abstract}

\noindent AMS Classification Primary 30G35; Secondary 15A69



\section{Introduction}
\setcounter{equation}{0}

In this paper we extend the concept of tensor product to the bicomplex case and prove the bicomplex counterpart of the classical Choi theorem in the theory of complex matrices and operators. We link these results to the theory of quantum channels in the bicomplex and hyperbolic case, as well as bicomplex DSP as in~\cite{AGS, EAI, GCT, GCT2}. In this introduction we give a short motivation on our choice to work in the bicomplex space.

In the real four dimensional case we have several models of writing a single variable theory and the two main ones are the bicomplex and the quaternionic model and, while a quaternionic number $q=x_1+x_2\, i+x_3\, j+x_4\, k$ and a bicomplex number $Z=x_1+x_2\i+x_3\j+x_4\k$ look similar, the fundamental difference is that in the anti-commutative quaternionic case $i\,j=-j\,i=k$, while in the commutative bicomplex case we have  $\i\,\j=\j\,\i=\k$.  These properties yield another imaginary unit $k=i \, j$ for the quaternions and a hyperbolic one $\k=\i \, \j$ in the bicomplex one.

There have been many attempts to understand a tensor product in the quaternionic case, however,  the attempts to define a probability measure that agrees with such a theory have, so far, been fruitless. For example, the tensor products that exist in literature do not preserve positivity of quaternionic matrices. In fact, quaternionic tensor products are base dependent and do not preserve many properties~\cite{Horwitz}, such as quaternionic positivity, for example. 
\begin{remark}
One can define two versions of a quaternionic tensor product of two quaternion matrices $Q_1=M_1+j\, N_1$ and $Q_2=M_2+j\, N_2$ as follows, the first as $Q_1 \otimes Q_2 =M_1\otimes M_2 +j\, N_1\otimes N_2 $ and the second $Q_1 \otimes Q_2 = M_1\otimes M_2 -N_1\otimes \bar{N_2} +j (M_1 \otimes  N_2 + N_1 \otimes \bar{M_2})$. 
Then, if we consider a quaternionic positive matrix to be one whose associated complex matrix 
$ \begin{pmatrix}
	M_1 && N_1\\
	-\bar{N_1} && \bar{M_1}
 \end{pmatrix}
$
is positive as a complex matrix, then neither of the quaternionic tensor products above will preserve positivity in this sense. 
In this case it is also difficult to define a probability measure.
\end{remark}

However, in the bicomplex case we can define a tensor product that suits notions of positivity as well as probabilistic ones well and many important theorems and applications in information and probability theory can be extended. Also, apart from the commutativity properties of the bicomplex algebra we have, in this case, a split along the $0-$divisors in the theory that enables such concepts to be defined and used with ease.
\smallskip

This paper adopts the following scheme: in Section~\ref{bicomplex} we review concepts of the algebra of bicomplex numbers and its hyperbolic subalgebra. In Section~\ref{Pos-bcplx-matr} we review concepts of positivity for the algebra of bicomplex matrices and prove some alternate statements. In Section~\ref{BC-tensor} we define the notion of bicomplex tensor product and we prove several properties; we also prove a result on recovering bicomplex components of the tensor product, as well as reduce all of the results of the section to the set of hyperbolic matrices. In Section~\ref{bcplx-choi} we prove a bicomplex counterpart of the complex Choi Theorem which shows that a bicomplex completely positive map is a finite bicomplex quantum channel.


\section{Preliminaries: Bicomplex Numbers}
\label{bicomplex}
\setcounter{equation}{0}

The algebra of bicomplex numbers was first introduced by Segre in~\cite{Segre}. During the past decades, a few isolated works analyzed either the properties of bicomplex numbers, or the properties of holomorphic functions defined on bicomplex numbers, and, without pretense of completeness, we direct the attention of the reader first to the to book of Price,~\cite{Price}, where a full foundation of the theory of multicomplex numbers was given, then to some of the works describing some analytic properties of functions in the field~\cite{alss, CSVV, DSVV, mltcplx}. Applications of bicomplex (and other hypercomplex) numbers can be also found in the works of Alfsmann, Sangwine, Gl\"{o}cker, and Ell~\cite{AGS}.\\

We now recall, in the same fashion as~\cite{CSVV,ESSV,Price}, the key definitions and results for the case of holomorphic functions of complex variables. The algebra of bicomplex numbers is generated by two commuting imaginary units $\i$
and $\j$ and we will denote the bicomplex space by $\BC$.  The product of the two commuting units  $\i$ and $\j$ is denoted by $ \k := \i\j$ and we note that $\k$ is a hyperbolic unit, i.e. it is a unit which squares to $1$.  Because of these various units in $\BC$, there are several
different conjugations that can be defined naturally. We will make use of these appropriate conjugations in this paper, and we refer the reader to~\cite{ESSV,mltcplx} for more information on bicomplex and multicomplex analysis.

\medskip


\subsection{Properties of the bicomplex algebra}
\label{bc}
The bicomplex space, $\BC$, is not a division algebra, and it has two distinguished zero
divisors, $\e_1$ and $\e_2$, which are idempotent, linearly independent, and mutually annihilating with respect to the
bicomplex multiplication:
\begin{align*}
  \e_1&:=\frac{1+\k}{2}\,,\qquad \e_2:=\frac{1-\k}{2}\,,\\
  \e_1 \cdot \e_2 &= 0,\qquad
  \e_1^2=\e_1 , \qquad \e_2^2 =\e_2\,,\\
  \e_1 +\e_2 &=1, \qquad \e_1 -\e_2 = \k\,.
\end{align*}
Just like $\{1,\mathbf{j} \},$ they form a basis of the complex algebra
$\BC$, which is called the {\em idempotent basis}. If we define the
following complex variables in $\C(\i)$:
\begin{align*}
  \beta_1 := z_1-\i z_2,\qquad \beta_2 := z_1+\i z_2\,,
\end{align*}
the $\C(\i)$--{\em idempotent representation} for $Z=z_1+\j z_2$ is
given by
\begin{align*}
  Z &= \beta_1\e_1+\beta_2\e_2\,.
\end{align*}

The $\C(\i)$--idempotent is the only representation for which
multiplication is component-wise, as shown in the next lemma.

\begin{remark}
  \label{prop:idempotent}
  The addition and multiplication of bicomplex numbers can be realized
  component-wise in the idempotent representation above. Specifically,
  if $Z= a_1\,\e_2 + a_2\,\e_2$ and $W= b_1\,\e_1 + b_2\,\e_2 $ are two
  bicomplex numbers, where $a_1,a_2,b_1,b_2\in\C(\i)$, then
  \begin{eqnarray*}
    Z+W &=& (a_1+b_1)\,\e_1  + (a_2+b_2)\,\e_2   ,  \\
    Z\cdot W &=& (a_1b_1)\,\e_1  + (a_2b_2)\,\e_2   ,  \\
    Z^n &=& a_1^n \,\e_1  + a_2^n \,\e_2  .
  \end{eqnarray*}
  Moreover, the inverse of an invertible bicomplex number
  $Z=a_1\e_1 + a_2\e_2 $ (in this case $a_1 \cdot a_2 \neq 0$) is given
  by
  $$
  Z^{-1}= a_1^{-1}\e_1 + a_2^{-1}\,\e_2 ,
  $$
  where $a_1^{-1}$ and $a_2^{-1}$ are the complex multiplicative
  inverses of $a_1$ and $a_2$, respectively.
\end{remark}

One can see this also by computing directly which product on the
bicomplex numbers of the form
\begin{align*}
  x_1 + \i x_2 + \j x_3 + \k x_4,\qquad x_1,x_2,x_3,x_4\in\R
\end{align*}
is component-wise, and one finds that the only one with this property
is given by the mapping:
\begin{align}
  \label{shakira}
  x_1 + \i x_2 + \j x_3 + \k x_4 \mapsto ((x_1 + x_4) + \i (x_2-x_3), (x_1-x_4) + \i (x_2+x_3))\,,
\end{align}
which corresponds exactly with the idempotent decomposition
\begin{align*}
  Z = z_1 + \j z_2 = (z_1- \i z_2)\e_1 + (z_1+ \i z_2)\e_2\,,
\end{align*}
where $z_1 = x_1+ \i x_2$ and $z_2 = x_3+ \i x_4$.

\begin{remark}
The idempotent representation splits the bicomplex space in $\BC=\mathbb C \mathbf{e}_1\bigoplus \mathbb C \mathbf{e}_2$, as:
\begin{equation}
  Z=z_1+\j z_2=(z_1-\i z_2)\mathbf{e}_1+(z_1+\i z_2)\mathbf{e}_2=\lambda_1\e_1+\lambda_2\e_2.
\end{equation}
\end{remark}

Simple algebra yields:
\begin{equation}
\begin{split}
  z_1&=\frac{\lambda_1+\lambda_2}{2}\\
    z_2&=\frac{\i(\lambda_1-\lambda_2)}{2}.
  \end{split}
  \end{equation}

There are several different conjugations that can be defined naturally and we will now define the conjugates in the bicomplex setting, as in~\cite{CSVV,ESSV}

\begin{definition} For any $Z\in \BC$ we have the following three conjugates:
  \begin{eqnarray}
  \label{conj}
    \overline{Z}=\overline{z_1}+\j\overline{z_2}\\
     Z^{\dagger}=z_1-\j z_2\\
      Z^*=\overline{Z^{\dagger}}=\overline{z_1}-\j\overline{z_2}.
  \end{eqnarray}
\end{definition}

We refer the reader to~\cite{ESSV} for more details.

\bigskip

\subsection{Hyperbolic subalgebra and the hyperbolic-valued modulus}

A special subalgebra of $\BC$ is the set of hyperbolic numbers, denoted by $\D$.  This
algebra and the analysis of hyperbolic numbers have been studied, for
example, in~\cite{alss,ESSV,sobczyk} and we summarize
below only the notions relevant for our results.  A {\em hyperbolic number}
can be defined independently of $\BC$, by $\mathfrak{z}=x + \k y$,
with $x,y,\in\R$, $\k\not\in\R, \k^2=1$, and we denote by $\D$ the
algebra of hyperbolic numbers with the usual component--wise addition
and multiplication.  The hyperbolic {\em conjugate} of $\mathfrak z$
is defined by $\mathfrak{z}^\diamond := x - \k y$, and note that:
\begin{equation}
  \mathfrak{z}\cdot \mathfrak{z}^\diamond=x^2-y^2\in\R\,,
\end{equation}
which yields the notion of the square of the {\em modulus} of a
hyperbolic number $\mathfrak{z}$, defined by
$ |\mathfrak{z}|_{\D}^2:=\mathfrak{z}\cdot \mathfrak{z}^\diamond$.
\begin{remark}
It is worth noting that both $\overline{Z}$ and $Z^{\dagger}$ reduce to  $\mathfrak{z}^\diamond$ when $Z= \mathfrak{z}$. In particular $\e_2=\e_1^\diamond=\e_1^*=\e_1^{\dagger}$.
\end{remark}
Similar to the bicomplex case, hyperbolic numbers have a unique
idempotent representation with real coefficients:
\begin{align}
  \label{D_idempotent}
  \mathfrak{z}=s \e_1 + t \e_2 \,,
\end{align}
where, just as in the bicomplex case, $\displaystyle \e_1 = \frac{1}{2} \left( 1 + \k \right) $,
$\displaystyle \e_2= \frac{1}{2} \left( 1 - \k \right)$, and
$s:=x+y$ and $t:=x-y$. Note that $\e_1^\diamond=\e_2$ if we consider
$\D$ as a subset of $\BC$, as briefly explained in the remark above. We also observe that
$$
|\mathfrak{z}|_{\D}^2 = x^2 - y^2 = (x+y)(x-y) = st.
$$

The hyperbolic algebra $\D$ is a subalgebra of the bicomplex numbers
$\BC$ (see~\cite{ESSV} for details). Actually $\BC$ is the
algebraic closure of $\D$, and it can also be seen as the
complexification of $\D$ by using either of the imaginary unit $\i$ or
the unit $\j$.
\begin{definition}
Define the set $\D^+$ of {\em non--negative hyperbolic numbers} by:
\begin{align*}
  \D^+ &= \left\{ x + \k y \, \big| \, x^2 - y^2 \geq 0,  x \geq 0 \right\}
       = \left\{ x + \k y \, \big| \, x \geq 0,  | y | \leq x \right\} \\
       &= \{ s \e_1 + t \e_2 \, \big| \, s, t \geq 0 \}.
\end{align*}
\end{definition}
\begin{remark}
As studied extensively in~\cite{alss}, one can define a partial order
relation defined on $\D$ by:
\begin{align}
  \label{po}
  \mathfrak{z}_1 \preceq \mathfrak{z}_2\qquad\text{if and only if}\qquad
  \mathfrak{z}_2-\mathfrak{z}_1\in\D^+,
\end{align}
and we will use this partial order to study the
{\em hyperbolic--valued} norm, which was first introduced and studied
in~\cite{alss}.
\end{remark}

The Euclidean norm $\|Z\|$ on $\BC$, when it is seen as
$\C^2(\i), \C^2(\j)$ or $\R^4$ is:
\begin{align*}
  \|Z\| = \sqrt{ | z_1 | ^2 + | z_2 |^2 \, } = \sqrt{ \Re\left( | Z |_\k^2 \right) \, } = \sqrt{
  \, x_1^2 + y_1^2 + x_2^2 + y_2^2 \, }.
\end{align*}
As studied in detail in~\cite{ESSV}, in idempotent
coordinates $Z=\lambda_1\e_1+\lambda_2\e_2$, the Euclidean norm becomes:
\begin{align}
  \label{Euclidean_idempotent}
  \|Z\| = \frac{1}{\sqrt2}\sqrt{|\lambda_1|^2 + |\lambda_2|^2}.
\end{align}


It is easy to prove that
\begin{align}
  \|Z \cdot W\|  \leq  \sqrt{2} \left(\|Z\| \cdot  \|W\| \right),
\end{align}
and we note that this inequality is sharp since if $Z = W= \e_1$, one
has:
\begin{align*}
  \|\e_1 \cdot \e_1\| = \|\e_1\| = \frac{1}{\sqrt{ 2}} =  \sqrt{2}\, \|\e_1\| \cdot \|\e_1\|,
\end{align*}
and similarly for $\e_2$.

\begin{definition}
One can define a
{\em hyperbolic-valued} norm for $Z=z_1+\j z_2 = \lambda_1 \e_1+\lambda_2\e_2$
by:
\begin{align*}
  \| Z\|_{\D_+} := |\lambda_1|\e_1 + |\lambda_1|\e_2 \in\D^+.
\end{align*}
\end{definition}
It is shown in~\cite{alss} that this definition obeys the corresponding properties
of a norm, i.e. $ \| Z\|_{\D_+}=0$ if and only if $ Z=0$,
it is multiplicative, and it
respects the triangle inequality with respect to the order introduced
above.

The previous norm can be generalized to the space of $\BC$ vectors, i.e. elements of $\BC^n$, and we will also define an inner product on the space of vectors in $\BC$.
Let  $\displaystyle \langle X,\, Y \rangle$ be the usual Hermitian inner product on $\C^n$, then we have the following:

\begin{definition}
For any $\mathsf  X,\mathsf Y\in \BC^n$, we have the following $\D-$valued inner product
\begin{equation}
\label{ip_def_bc_n}
 \langle \mathsf  X,\, \mathsf  Y \rangle_{\D}= \langle X_1,\, Y_1 \rangle \e_1+ \langle X_2,\, Y_2 \rangle \e_2,
\end{equation}
where $\mathsf  X=X_1\e_1 +X_2\e_2$ and  $\mathsf  Y=Y_1\e_1 +Y_2\e_2$, and $X_l,\, Y_l\in \C^n$ for $l=1,2$.
\end{definition}

This inner product yields the hyperbolic-valued modulus of a vector  $\mathsf  X=X_1\e_1 +X_2\e_2$ as: $$\| \mathsf  X\|_{\D_+}=\|X_1\|\e_1 +\|X_2\|\e_2.$$


\section{Positive bicomplex matrices}
\label{Pos-bcplx-matr}
\setcounter{equation}{0}
In this section we review a characterization of positive bicomplex matrices and we start with the definition.

\begin{definition}[Positive $ \mathbb{BC}$ matrix]
A matrix $A=(a_{jl}) \in \mathbb{BC}^{n \times n}$ is hyperbolic positive if for all $ c =(c_1,..., c_n)^{t}\in \mathbb{BC}^{n \times 1}$ we have
\begin{equation}
(c^{*})^t  \cdot A \cdot c  \in \mathbb{D}^{+}.
\end{equation}
\end{definition}
\label{char}
From \cite[Proposition 2.2.7]{alss} we know the following characterization
\begin{proposition}
\label{equivalence}
Let
$$ A = A_1 + \mathbf{j}A_2 = \mathcal{A}_1\e_1 + \mathcal{A}_2\e_2,$$
be an element of $ \mathbb{BC}^{n \times n}$ with $A_1$, $A_2$, $ \mathcal{A}_1$ and $\mathcal{A}_2$ in $ \mathbb{C}^{n \times n}(\mathbf{j})$. Then, the following are equivalent 
\begin{itemize}
\item[1)] $A$ is hyperbolic positive,
\item[2)] both $ \mathcal{A}_1$ and $ \mathcal{A}_2$ are complex positive matrices,
\item[3)] $A_1 \geq 0$, the matrix $A_2$ is skew self adjoint, that is, $A_2 + \bar{A}_2^t=0$ and
$$ -A_1 \leq i A_2 \leq A_1,$$
\item[4)]  all eigenvalues are in $ \mathbb{D}^{+}$.
\end{itemize}
\end{proposition}
Other characterizations of hyperbolic positive matrices were proved in \cite{alss}, as follows.

\begin{proposition}
Let $A \in \mathbb{BC}^{n \times n}$. The following are equivalent
\begin{itemize}
\item[1)] A is hyperbolic positive,
\item[2)] $A=B^{*t}\cdot B$ where $B \in \mathbb{BC}^{m \times n}$ for some $m \in \mathbb{N}$,
\item[3)] $A=C^2$ where the matrix $C$ is hyperbolic positive,
\item[4)] $A$ is $*$ - Hermitian (i.e., $A=(A^{*})^{t}$).
\end{itemize}
\end{proposition}

See \cite[Prop. 2.2.8]{alss} and \cite[Prop. 2.2.9]{alss} for the proofs of the previous results.

In this paper we show a further characterization of hyperbolic positive matrices.

\begin{theorem}
Let $A \in \mathbb{BC}^{n \times n}$. Then the following conditions are equivalent
\begin{itemize}
\item[1)]  the matrix $A$ is hyperbolic positive,
\item[2)] there exists an upper triangular matrix $U \in \mathbb{BC}^{m \times n}$ such that $A= U^{*t} U$,
\item[3)] there exists a lower triangular matrix $L \in \mathbb{BC}^{m \times n}$ such that $A= L^{*t} L$, 
\item[4)]  there exist $a_1,..., a_r \in \mathbb{BC}^{n \times 1}$, that may be chosen pairwise orthogonal such that
$$ A= \sum_{i=1}^{r} a_i a_{i}^{*t}.$$
\end{itemize}
\end{theorem} 
\begin{proof}
Let $A \in \mathbb{BC}^{n \times n}$ be a matrix decomposed by its idempotent decomposition $ A= \mathcal{A}_1\e_1+ \mathcal{A}_2\e_2$, where $ \mathcal{A}_1$ and $ \mathcal{A}_2$ are matrices in $ \mathbb{C}^{n \times n}(\i)$. 

We start proving $1) \Longrightarrow 2)$. By Proposition \ref{equivalence} we know that $ \mathcal{A}_1$ and $ \mathcal{A}_2$ are complex positive matrices. Then by \cite[theorem 3.5.30]{A} we get that
$$ \mathcal{A}_1:= \overline{V}^tV \quad \hbox{and} \quad \mathcal{A}_2:= \overline{W}^tW,$$
where $U$ and $V$ are upper triangular matrices of $ \mathbb{C}^{n \times n}(\i)$. This implies $A=U^{*t} U$, with $U:=V\e_1+W\e_2$.

Now we show $2) \Longrightarrow 1)$. We write $U=V\e_1+W\e_2$, then we get
$$ A=\overline{V}^tV\e_1+\overline{W}^tW\e_2.$$
Since the matrices $\overline{V}^tV$ and $\overline{W}^tW$ are complex positive matrices by Proposition \ref{equivalence} we get that $A$ is an hyperbolic positive matrix.

By similar arguments it is possible to show the equivalence $1) \Longleftrightarrow 3).$

Now we show $1) \Longleftrightarrow 4)$ Since the matrices $ \mathcal{A}_1$ and $ \mathcal{A}_2$ are complex positive matrices by \cite[Thm. 3.5.30]{A} we get that
$$ \mathcal{A}_{1}= \sum_{i=1}^r b_i \bar{b}_i^t \qquad \mathcal{A}_{2}= \sum_{i=1}^r c_i \bar{c}_i^t,$$
where $b_1,..., b_r \in \mathbb{C}^{n \times 1}(\i)$ are pairwise orthogonal and $c_1,..., c_r \in \mathbb{C}^{n \times 1}(\i)$ are pairwise orthogonal. This implies that 
$$ A= \sum_{i=1}^{r} a_i a_{i}^{*t},$$
where $a_i:= b_i \e_1+c_i\e_2$.

Finally we show $4) \Longleftrightarrow 1)$. By writing $a_i= b_i \e_1+c_i\e_2$ we can write
$$A=\sum_{i=1}^r b_i \bar{b}_i^t \e_1+\sum_{i=1}^r c_i \bar{c}_i^t\e_2.$$
By \cite[theorem 3.5.30]{A} we get that the matrices $\sum_{i=1}^r b_i \bar{b}_i^t$ and $\sum_{i=1}^r c_i \bar{c}_i^t$ are positive and hence by Proposition \ref{equivalence} we get that the matrix $A$ is hyperbolic positive.
\end{proof}

\begin{definition}
	A matrix $A \in \mathbb{BC}^{n \times n}$ is called state if it is hyperbolic positive and if $ \hbox{Tr}(A) =1$. 
\end{definition}

The following theorem characterizes bicomplex states:
\begin{theorem}
 A matrix $A=\mathcal{A}_1 \e_1+ \mathcal{A}_2 \e_2$, with $\mathcal{A}_1$, $\mathcal{A}_2 \in \mathbb{C}^{n \times n}(\i)$, is a state if and only if $\mathcal{A}_1$ and $ \mathcal{A}_2$ are state matrices.
\end{theorem}
\begin{proof}
	By Theorem~\ref{char1} it is enough to show that $ \hbox{Tr}(A)=1$ if and only if $\hbox{Tr}(\mathcal{A}_1)=\hbox{Tr}(\mathcal{A}_2)=1$.
	Since the trace is a linear operator we get
	\begin{equation}
		\label{tra}
		\hbox{Tr}(A)= \hbox{Tr}(\mathcal{A}_1)\e_1+ \hbox{Tr}(\mathcal{A}_2) \e_2.
	\end{equation}
	If $\hbox{Tr}(\mathcal{A}_1)=\hbox{Tr}(\mathcal{A}_2)=1$, form \eqref{tra} we get
	\begin{eqnarray*}
		\hbox{Tr}(A)&=& \e_1+ \e_2\\
		&=& 1.
	\end{eqnarray*}
	Now, let us suppose that $\hbox{Tr}(A)=1$ then form \eqref{tra} we obtain
	\begin{equation}
		\label{tra1}
		\hbox{Tr}(\mathcal{A}_1)\e_1+ \hbox{Tr}(\mathcal{A}_2) \e_2=1.
	\end{equation}
	However, we know that $1=\e_1+\e_2$ and so we can rewrite formula \eqref{tra1} as
	\begin{equation}
		\label{tra2}
		\hbox{Tr}(\mathcal{A}_1)\e_1+ \hbox{Tr}(\mathcal{A}_2) \e_2=\e_1+\e_2.
	\end{equation}
	Firstly, we multiply formula \eqref{tra2} by $\e_1$, since $\e_1^2=1$ we get that
	$$ \hbox{Tr}(\mathcal{A}_1)=1,$$ 
	Now, we multiply formula \eqref{tra2} by $\e_2$, since $\e_2^2=1$ we obtain that
	$$\hbox{Tr}(\mathcal{A}_2)=1.$$
	This proves the first part of the statement.
\end{proof}

Now, we define the inverse of a matrix with bicomplex entries
\begin{definition}
\label{inve}
The inverse of a bicomplex matrix $A \in \mathbb{BC}^{n \times n}$ is the bicomplex matrix denoted by $A^{-1}$ such that
$$ AA^{-1}=A^{-1}A= I_n,$$
where $I_n$ is the $n \times n$ identity matrix. 
\end{definition}
\begin{lemma}
\label{inve3}
Let $A= \mathcal{A}_1 \e_1+ \mathcal{A}_2 \e_2 \in \mathbb{BC}^{n \times n}$, where $ \mathcal{A}_1$, $ \mathcal{A}_2 \in \mathbb{C}^{n \times n}(i)$. We can write its inverse matrix as
$$ A^{-1}= \mathcal{A}_1^{-1} \e_1+ \mathcal{A}_2^{-1} \e_2.$$
\end{lemma}
\begin{proof}
By Definition \ref{inve} we have to show that $AA^{-1}=A^{-1}A=I_n.$
By the fact thats $\e_1+\e_2=1$ and $\e_1\e_2=0$ we get
\begin{eqnarray*}
A^{-1}A&=&(\mathcal{A}_1^{-1} \e_1+ \mathcal{A}_2^{-1} \e_2)(\mathcal{A}_1 \e_1+ \mathcal{A}_2 \e_2 )\\
&=& (I_n )\e_1+ (I_n) \e_2\\
&=& I_n.
\end{eqnarray*}
The equality $AA^{-1}=I_n$ follows by similar computations.
\end{proof}
Now, we show that we can write an inverse matrix in a different way
\begin{theorem}
Let $A=A_1+A_2\textbf{j}$ with $A_1$, $A_2 \in \mathbb{C}^{n \times n}(\textbf{i})$. Then we have
\begin{equation}
\label{inve2}
A^{-1}:= \tilde{A}_1+\tilde{A}_2\textbf{j},
\end{equation}
where $$ \tilde{A}_1:= \frac{1}{2}[\mathcal{A}_1^{-1}+ \mathcal{A}_2^{-1}] \quad  \hbox{and}  \quad \tilde{A}_2:= -\frac{i}{2}[\mathcal{A}_2^{-1}-\mathcal{A}_1^{-1}],$$
with $ \mathcal{A}_1$, $ \mathcal{A}_2 \in \mathbb{C}^{n \times n}(i)$.
\end{theorem}
\begin{proof}
By using the fact that $\e_1= \frac{1+ \textbf{ij}}{2}$ and $\e_1= \frac{1- \textbf{ij}}{2}$ we get
\begin{eqnarray*}
A^{-1}&=&\frac{1}{2}[\mathcal{A}_1^{-1}+ \mathcal{A}_2^{-1}] - \frac{\textbf{i} \textbf{j}}{2}[\mathcal{A}_2^{-1}-\mathcal{A}_1^{-1}]\\
&=& \left( \frac{1+ \textbf{i} \textbf{j}}{2}\right)\mathcal{A}_1^{-1}+ \left( \frac{1- \textbf{i} \textbf{j}}{2}\right)\mathcal{A}_2^{-1}\\
&=&  \mathcal{A}_1^{-1}\e_1+  \mathcal{A}_2^{-1}\e_2.
\end{eqnarray*}
We get the thesis by Lemma \ref{inve3}.
\end{proof}
 

\section{Tensor product of bicomplex matrices}
\label{BC-tensor}
\setcounter{equation}{0}

In this section we first write a coherent definition of the bicomplex tensor product, reduce it to the hyperbolic subspace as well, then we follow by a method of finding the components of the bicomplex tensor from the outcome matrix, ending with an application to digital signal processing.


\subsection{Definition and Properties of the Bicomplex Tensor Product}

We now turn to the definition of the bicomplex tensor product of two matrices.
\begin{definition}
	\label{tensor}
	Let us consider $A=A_1 + \mathbf{j}A_2\, \in \mathbb{BC}^{n_1\times n_2}$ and $B=B_1 + \mathbf{j}B_2\, \in \mathbb{BC}^{m_1\times m_2}$ with $A_1,\, A_2 \in \mathbb{C}^{n_1 \times n_2}(\i)$, $ B_1,\, B_2 \in \mathbb{C}^{m_1 \times m_2}(\i)$. Then we define the bicomplex tensor product of these two matrices as
	$$ A \otimes_{\mathbf{j}} B=(A_1 \otimes B_1- A_2 \otimes B_2)+ \mathbf{j}(A_1 \otimes B_2 +A_2 \otimes B_1),$$
	where $ A \otimes_{\mathbf{j}} B \, \in \mathbb{BC}^{n_1 m_1\times n_2 m_2}$
\end{definition}
\begin{proposition}
	Let us consider $A=A_1 + \mathbf{j}A_2$ and $B=B_1 + \mathbf{j}B_2$ with $ A_1$, $A_2$, $ B_1$, $ B_2 $ as in Definition~\ref{tensor}. Then we have
	$$ \begin{pmatrix}
		A_1 && -A_2\\
		A_2 && A_1
	\end{pmatrix} \otimes_{\mathbf{j}}
	\begin{pmatrix}
		B_1 && -B_2\\
		B_2 && B_1
	\end{pmatrix}= \begin{pmatrix}
		A_1 \otimes B_1- A_2 \otimes B_2 && -A_1 \otimes B_2 -A_2 \otimes B_1\\
		A_1 \otimes B_2 +A_2 \otimes B_1 && A_1 \otimes B_1- A_2 \otimes B_2
	\end{pmatrix}.$$
\end{proposition}
\begin{proof}
	It follows from Definition \ref{tensor} and the matrix representation of the bicomplex setting.
\end{proof}

\begin{example}
\label{example1}
Let $A= \begin{pmatrix}
1 & 1+\i\\
0 & \i
\end{pmatrix}+\mathbf{j}\begin{pmatrix}
\i & 1\\
1 & \i
\end{pmatrix}$
and
$B= \begin{pmatrix}
	0 & \i\\
	\i & 1
\end{pmatrix}+\mathbf{j}\begin{pmatrix}
	1 & 0\\
	1 & \i+1
\end{pmatrix}$
be matrices in $ \mathbb{BC}^{2 \times 2}$. Then by Definition \ref{tensor} we have
\begin{equation}
\label{ex2}
	A \otimes_{\mathbf{j}} B= \begin{pmatrix}
		-\i & \i & -1 & \i-1\\
		0 & 2-\i & \i-2 & 0\\
		-1 & 0 & - \i &-1\\
		-1 & -\i-1& -\i-1& 1
	\end{pmatrix}
	+\mathbf{j}
	\begin{pmatrix}
		1 & -1& \i+1& \i\\
		0 & 2\i+1& 2\i+1& 2\i+1\\
		0 & \i& \i& -1\\
		\i & 1 & \i-1 & 2\i-1
	\end{pmatrix}
\end{equation}
\end{example}

Now, we show that we can write the bicomplex tensor product in an equivalent way
\begin{theorem}
	\label{char1}
	Let $A= \mathcal{A}_1 \e_1+ \mathcal{A}_2 \e_2 \in \mathbb{BC}^{n_1 \times n_2}$, $ B= \mathcal{B}_1 \e_1+ \mathcal{B}_2 \e_2  \in \mathbb{BC}^{m_1 \times m_2}$ with $\mathcal{A}_1$, $\mathcal{A}_2 \in \mathbb{C}^{n_1 \times n_2}(\i)$ and $\mathcal{B}_1$, $\mathcal{B}_2 \in \mathbb{C}^{m_2 \times m_2}(\i)$. Then we have
	$$A \otimes_{\mathbf{j}} B=(\mathcal{A}_1 \otimes \mathcal{B}_1) \e_1+ (\mathcal{A}_2 \otimes \mathcal{B}_2) \e_2,$$
with $ A \otimes B \in \mathbb{BC}^{m_1n_1 \times m_2n_2}$.
\end{theorem}
\begin{proof}
	Firstly, by recalling that $\e_1= \frac{1+\i \mathbf{j}}{2}$ and $\e_2= \frac{1-\i\mathbf{j}}{2}$ we get
	\begin{eqnarray*}
		A &=& \mathcal{A}_1 \e_1+ \mathcal{A}_2 \e_2\\
		&=& \frac{1}{2} (\mathcal{A}_1+\mathcal{A}_2)- \frac{\mathbf{j}\i}{2}(\mathcal{A}_2-\mathcal{A}_1)\\
		&=:& A_1+ \mathbf{j}A_2, 
	\end{eqnarray*}	
	where 
	\begin{equation}
		\label{ma}
		A_1= \frac{1}{2} (\mathcal{A}_1+ \mathcal{A}_2) \qquad A_2=- \frac{\i}{2}(\mathcal{A}_2-\mathcal{A}_1).
	\end{equation}
	Similarly, for the matrix $B$ we have
	\begin{eqnarray*}
		B&=& \mathcal{B}_1 \e_1+ \mathcal{B}_2 \e_2\\
		&=:& B_1+ \mathbf{j}B_2,
	\end{eqnarray*}
	where 
	\begin{equation}
		\label{mb}
		B_1= \frac{1}{2} (\mathcal{B}_1+ \mathcal{B}_2) \qquad B_2=- \frac{\i}{2}(\mathcal{B}_2-\mathcal{B}_1).
	\end{equation}
	Let us denote
	\begin{equation}
		\label{c1}
		C_1:=A_1 \otimes B_1- A_2 \otimes B_2,
	\end{equation}
	and
	\begin{equation}
		\label{c2}
		C_2:= A_1 \otimes B_2 +A_2 \otimes B_1.
	\end{equation}
	Now, we substitute the expressions \eqref{ma}, \eqref{mb} in \eqref{c1} and \eqref{c2}
	\begin{eqnarray*}
		C_1&=&A_1 \otimes B_1- A_2 \otimes B_2\\
		&=& \frac{1}{4} \biggl(\mathcal{A}_1 \otimes \mathcal{B}_1 + \mathcal{A}_1 \otimes \mathcal{B}_2+ \mathcal{A}_2 \otimes \mathcal{B}_1+ \mathcal{A}_2 \otimes \mathcal{B}_2\biggr)+\\
		&& + \frac{1}{4} \biggl(\mathcal{A}_2 \otimes \mathcal{B}_2 -\mathcal{A}_2 \otimes \mathcal{B}_1-\mathcal{A}_1 \otimes \mathcal{B}_2 + \mathcal{A}_1 \otimes \mathcal{B}_1 \biggr)\\
		&=& \frac{1}{2} \biggl( \mathcal{A}_1 \otimes \mathcal{B}_1+ \mathcal{A}_2 \otimes \mathcal{B}_2\biggr)
	\end{eqnarray*}
	and 
	\begin{eqnarray*}
		C_2&=&A_1 \otimes B_2+ A_2 \otimes B_1\\
		&=& -\frac{\i}{4} \biggl(\mathcal{A}_1 \otimes \mathcal{B}_2 - \mathcal{A}_1 \otimes \mathcal{B}_1+ \mathcal{A}_2 \otimes \mathcal{B}_2- \mathcal{A}_2 \otimes \mathcal{B}_1\biggr)+\\
		&& -\frac{\i}{4} \biggl(\mathcal{A}_2 \otimes \mathcal{B}_1 +\mathcal{A}_2 \otimes \mathcal{B}_2-\mathcal{A}_1 \otimes \mathcal{B}_1 - \mathcal{A}_1 \otimes \mathcal{B}_2 \biggr)\\
		&=& -\frac{\i}{2} \biggl( \mathcal{A}_2 \otimes \mathcal{B}_2- \mathcal{A}_1 \otimes \mathcal{B}_1\biggr)
	\end{eqnarray*}
	By definition \ref{tensor} we get
	\begin{eqnarray*}
		A \otimes_{\mathbf{j}} B&=& C_1 +\j C_2\\
		&=&\frac{1}{2} \biggl( \mathcal{A}_1 \otimes \mathcal{B}_1+ \mathcal{A}_2 \otimes \mathcal{B}_2\biggr)-\frac{\i\j }{2} \biggl( \mathcal{A}_2 \otimes \mathcal{B}_2- \mathcal{A}_1 \otimes \mathcal{B}_1\biggr)\\
		&=& (\mathcal{A}_1 \otimes \mathcal{B}_1) \e_1+ (\mathcal{A}_2 \otimes \mathcal{B}_2) \e_2.
	\end{eqnarray*}
\end{proof}
\begin{example}
Let us consider the matrices $A$ and $B$ like in the example \ref{example1}. We split the matrices in the respective idempotent decomposition
$$ A= \begin{pmatrix}
2 &1 \\
-\i & \i+1
\end{pmatrix} \e_1+ \begin{pmatrix}
0 & 2 \i+1\\
\i & \i-1
\end{pmatrix} \e_2 $$
and
$$ B= \begin{pmatrix}
-\i & \i\\
0 & 2-\i
\end{pmatrix} \e_1+ \begin{pmatrix}
\i & \i\\
2\i & \i
\end{pmatrix} \e_2
$$
Then by theorem \ref{char1} we have
\begin{equation}
\label{ex3}
A \otimes_{\mathbf{j}} B= \begin{pmatrix}
-2 \i & 2 \i & - \i& \i\\
0 & 4-2\i& 0 & 2-\i\\
-1& 1& 1-\i& \i-1\\
0 & -2\i-1 & 0 & \i+3
\end{pmatrix}\e_1
+\begin{pmatrix}
0 & 0 & \i-2& \i-2\\
0 & 0& 2\i-4& \i-2\\
-1 & -1& -1-\i&-1-\i\\
-2 & -1& -2-2\i& -1-\i
\end{pmatrix}\e_2
\end{equation}
Formula \eqref{ex3} is equal to \eqref{ex2}.
\end{example}

We now write the fundamental properties of the bicomplex tensor product and we see that we recover all properties, including ones that characterize the hyperbolic positivity of a tensor product of two hyperbolic positive matrices.
\begin{theorem}
Let $A, B \in \mathbb{BC}^{n \times n}$ be two hyperbolic positive matrices, then $ A \otimes_{\mathbf{j}} B$ is a hyperbolic positive matrix. Moreover, if $A, B \in \mathbb{BC}^{n \times n}$ are states (i.e. hyperbolic positive matrices of trace $1$), then $A \otimes_{\mathbf{j}} B$ is also a bicomplex state. 
\end{theorem} 
\begin{proof}
	To prove the first part of this theorem, we split the matrices $A$ and $B$ in the idempotent decomposition
	$$A= \mathcal{A}_1 \e_1+ \mathcal{A}_2 \e_2, \qquad B= \mathcal{B}_1 \e_1+ \mathcal{B}_2 \e_2, $$ with $\mathcal{A}_1$, $\mathcal{A}_2$, $\mathcal{B}_1$, $\mathcal{B}_2 \in \mathbb{C}^{n \times n}(\i)$. By Theorem~\ref{char1} we know that $A$ and $B$ are hyperbolic positive if and only if  $\mathcal{A}_1$, $\mathcal{A}_2$, $\mathcal{B}_1$, $\mathcal{B}_2$ are complex positive matrices. This implies that
	$$ \mathcal{A}_{1} \otimes \mathcal{A}_2 \geq 0,$$
	and
	$$ \mathcal{B}_{1} \otimes \mathcal{B}_2 \geq 0.$$
	Finally, by Proposition~\ref{equivalence} we get that $A \otimes_{\mathbf{j}} B \geq 0.$
To prove the second part of the theorem, we use the fact the trace is a linear operator we get
	\begin{equation}
		\label{tra5}
		\hbox{Tr}(A \otimes_{\mathbf{j}} B)= \hbox{Tr}(\mathcal{A}_1 \otimes \mathcal{B}_1) \e_1+ \hbox{Tr}(\mathcal{A}_2 \otimes \mathcal{B}_2) \e_2.
	\end{equation}
	Since the matrices $\mathcal{A}_1$, $\mathcal{A}_2$, $\mathcal{B}_1$, $\mathcal{B}_2$ are complex matrices we know that 
	\begin{equation}
		\label{tra4}
		\hbox{Tr}(\mathcal{A}_l \otimes \mathcal{B}_l)=\hbox{Tr}(\mathcal{A}_l) \hbox{Tr}(\mathcal{B}_l),
	\end{equation}
	with $l=1,2$. By hypothesis $A$ and $B$ are state matrices, and so by theorem \ref{char1} we get that $ \mathcal{A}_l$ and $ \mathcal{B}_l$, with $l=1,2$, are positive state matrices. In particular this implies that
	\begin{equation}
		\label{tra3}
		\hbox{Tr}( \mathcal{A}_l)= \hbox{Tr}( \mathcal{B}_l)=1, \qquad l=1,2.
	\end{equation}
	By putting together \eqref{tra5} and \eqref{tra4},\eqref{tra3} we obtain 
	\begin{eqnarray*}
		\hbox{Tr}(A \otimes_{\mathbf{j}} B)&=& \hbox{Tr}(\mathcal{A}_1 \otimes \mathcal{B}_1) \e_1+ \hbox{Tr}(\mathcal{A}_2 \otimes \mathcal{B}_2) \e_2.\\
		&=&\hbox{Tr}(\mathcal{A}_1) \hbox{Tr}(\mathcal{B}_1)\e_1 + \hbox{Tr}(\mathcal{A}_2) \hbox{Tr}(\mathcal{B}_2)\e_2\\
		&=& \e_1+\e_2\\
		&=&1.
	\end{eqnarray*}
\end{proof}

\begin{theorem} The bicomplex tensor product has the following properties:
\begin{itemize}
\item[1)] it is bilinear, i.e. for $A$, $D$ in $\mathbb{BC}^{n_2 \times n_2}$  and $B$, $C$ in $ \mathbb{BC}^{m_1 \times m_2}$ we have
\begin{equation*}
A \otimes_{\mathbf{j}} (B+C)= A \otimes_{\mathbf{j}} B+A \otimes_{\mathbf{j}} C
\end{equation*}
and
\begin{equation*}
(A+D) \otimes_{\mathbf{j}} B=A \otimes_{\mathbf{j}} B+ D \otimes_{\mathbf{j}} B
\end{equation*}
\item[2)] it has the fundamental property of tensor products, i.e. for $A \in \mathbb{BC}^{n \times m}$, $B \in \mathbb{BC}^{m \times r}$, $C \in \mathbb{BC}^{\ell \times p}$ and $D \in \mathbb{BC}^{p \times q}$ we have
\begin{equation*}
AB \otimes_{\mathbf{j}} CD= (A \otimes_{\mathbf{j}} C)(B \otimes_{\mathbf{j}} D).
\end{equation*} 

\end{itemize}
\end{theorem}
\begin{proof}
For {1)} let us start by proving the first bilinear property (the second follows by similar arguments). We split the matrices $A$, $B$ and $C$ in the idempotent decomposition.
$$A= \mathcal{A}_1 \e_1+ \mathcal{A}_2 \e_2, \qquad B= \mathcal{B}_1 \e_1+ \mathcal{B}_2 \e_2, \qquad C= \mathcal{C}_1 \e_1+ \mathcal{C}_2 \e_2,$$ 
with $\mathcal{A}_1$, $\mathcal{A}_2 \in \mathbb{C}^{n \times n}(\i)$ and  $\mathcal{B}_1$, $\mathcal{B}_2 $, $\mathcal{C}_1$, $\mathcal{C}_2\in \mathbb{C}^{m \times m}(\i)$. Then by theorem \ref{char1} we get
\begin{eqnarray*}
A \otimes_{\mathbf{j}} (B+C)&=&(\mathcal{A}_1 \e_1+ \mathcal{A}_2 \e_2) \otimes_{\mathbf{j}}\left[(\mathcal{B}_1+\mathcal{C}_1) \e_1+(\mathcal{B}_2+\mathcal{C}_2) \e_2\right]\\
&=& [\mathcal{A}_1 \otimes (\mathcal{B}_1+ \mathcal{C}_1)]\e_1+[\mathcal{A}_2 \otimes (\mathcal{B}_2+ \mathcal{C}_2)]\e_2
\end{eqnarray*}
Since the matrices $ \mathcal{A}_s$, $ \mathcal{B}_s$ and $ \mathcal{C}_s$ have complex entries by \cite{A} we have
\begin{equation}
\label{bline3}
A \otimes_{\mathbf{j}} (B+C)=\left(\mathcal{A}_1 \otimes \mathcal{B}_1+ \mathcal{A}_1 \otimes \mathcal{C}_1 \right) \e_1+\left(\mathcal{A}_2 \otimes \mathcal{B}_2+ \mathcal{A}_2 \otimes\mathcal{C}_2 \right) \e_2.
\end{equation}
By theorem \ref{char1} we get
\begin{eqnarray}
\nonumber
A \otimes_{\mathbf{j}} B+ A \otimes_{\mathbf{j}} C&=& \left[(\mathcal{A}_1 \otimes \mathcal{B}_1)\e_1+(\mathcal{A}_2 \otimes \mathcal{B}_2)\e_2 \right]+\left[(\mathcal{A}_1 \otimes \mathcal{C}_1)\e_1+(\mathcal{A}_2 \otimes \mathcal{C}_2)\e_2 \right]\\
\label{bline4}
&=& \left(\mathcal{A}_1 \otimes \mathcal{B}_1+ \mathcal{A}_1 \otimes \mathcal{C}_1 \right) \e_1+\left(\mathcal{A}_2 \otimes \mathcal{B}_2+ \mathcal{A}_2 \otimes\mathcal{C}_2 \right) \e_2.
\end{eqnarray}
We get the thesis by the fact that \eqref{bline3} and \eqref{bline4} are equal.

To prove the fundamental identity {2)}, we start by splitting the matrices $A$ and $B$ by the idempotent decomposition
$$ A= \mathcal{A}_1 \e_1+ \mathcal{A}_2 \e_2, \qquad B= \mathcal{B}_1 \e_1+\mathcal{B}_2 \e_2,$$
with $\mathcal{A}_1$, $\mathcal{A}_2 \in \mathbb{C}^{n \times m}(\i)$ and $\mathcal{B}_1$, $\mathcal{B}_2 \in \mathbb{C}^{m \times r}(\i)$. Then
$$ AB= (\mathcal{A}_1 \mathcal{B}_1) \e_1+ (\mathcal{A}_2 \mathcal{B}_2) \e_2.$$
Similarly, for the matrices $C$ and $D$ we have
$$ C= \mathcal{C}_1 \e_1+ \mathcal{C}_2 \e_2, \qquad D= \mathcal{D}_1 \e_1+\mathcal{D}_2 \e_2,$$
with $\mathcal{C}_1$, $\mathcal{C}_2 \in \mathbb{C}^{\ell \times p}(\i)$, $\mathcal{D}_1$, $\mathcal{D}_2 \in \mathbb{C}^{p \times q}(\i)$. Then
$$ CD= (\mathcal{C}_1 \mathcal{D}_1) \e_1+ (\mathcal{C}_2 \mathcal{D}_2) \e_2.$$
Now, by theorem \ref{char1} we get
$$ (AB) \otimes_{\mathbf{j}} (CD)=(\mathcal{A}_1 \mathcal{B}_1  \otimes \mathcal{C}_1 \mathcal{D}_1) \e_1 +(\mathcal{A}_2 \mathcal{B}_2  \otimes \mathcal{C}_2 \mathcal{D}_2) \e_2.$$
Since the matrices $ \mathcal{A}_s$, $ \mathcal{B}_s$, $ \mathcal{C}_s$ and $ \mathcal{D}_s$ with $ s=1,2$ are matrices with complex entries by \cite{A} we obtain
\begin{equation}
\label{first}
 (AB) \otimes_{\mathbf{j}} (CD)=(\mathcal{A}_1   \otimes \mathcal{C}_1)(\mathcal{B}_1 \otimes \mathcal{D}_1) \e_1 +(\mathcal{A}_2   \otimes \mathcal{C}_2)(\mathcal{B}_2 \otimes \mathcal{D}_2) \e_2.
\end{equation}
On the other side, by theorem \ref{char1} we have
$$ A \otimes_{\mathbf{j}} C= (\mathcal{A}_1 \otimes \mathcal{C}_1) \e_1 +(\mathcal{A}_2 \otimes \mathcal{C}_2) \e_2,$$
and
$$ B \otimes_{\mathbf{j}} D= (\mathcal{B}_1 \otimes \mathcal{D}_1) \e_1 +(\mathcal{B}_2 \otimes \mathcal{D}_2) \e_2.$$
Hence, we get
\begin{equation}
\label{second}
( A \otimes_{\mathbf{j}} C)(B \otimes_{\mathbf{j}} D)=(\mathcal{A}_1   \otimes \mathcal{C}_1)(\mathcal{B}_1 \otimes \mathcal{D}_1) \e_1 +(\mathcal{A}_2   \otimes \mathcal{C}_2)(\mathcal{B}_2 \otimes \mathcal{D}_2) \e_2.
\end{equation}
Finally, since the right hand side of the equations \eqref{first} and \eqref{second} are equal we obtain the corresponding results.
\end{proof}

\begin{lemma}
Let $A \in \mathbb{BC}^{n \times n}$ and $ B \in \mathbb{BC}^{m \times m}$. Then we have
$$ (A \otimes_{\mathbf{j}} B)^{-1}= A^{-1} \otimes_{\mathbf{j}} B^{-1}.$$
\end{lemma} 
\begin{proof}
Let us start by writing the matrices $A$ and $B$ by the idempotent decomposition:
$$ A= \mathcal{A}_1 \e_1+ \mathcal{A}_2 \e_2 \qquad B= \mathcal{B}_1 \e_1+ \mathcal{B}_2 \e_2,$$
with $\mathcal{A}_1$, $\mathcal{A}_2 \in \mathbb{C}^{n \times n}(\i)$ and $\mathcal{B}_1$, $\mathcal{B}_2 \in \mathbb{C}^{m \times m}(\i)$. By theorem \ref{char1} we get
$$ A \otimes_{\mathbf{j}} B=(\mathcal{A}_1 \otimes \mathcal{B}_1) \e_1+(\mathcal{A}_2 \otimes \mathcal{B}_2) \e_2.$$
Then by Lemma \ref{inve3} we obtain
$$ (A \otimes_{\mathbf{j}} B)^{-1}=(\mathcal{A}_1 \otimes \mathcal{B}_1)^{-1} \e_1+(\mathcal{A}_2 \otimes \mathcal{B}_2)^{-1} \e_2.$$
Since the matrices $ \mathcal{A}_s$, $\mathcal{B}_s$, with $s=1,2$, have complex entries by \cite{A} we can write
\begin{equation}
\label{first1}
(A \otimes_{\mathbf{j}} B)^{-1}=(\mathcal{A}_1^{-1} \otimes \mathcal{B}_1^{-1}) \e_1+(\mathcal{A}_2^{-1} \otimes \mathcal{B}_2^{-1})^{-1} \e_2.
\end{equation}
Now, by Lemma \ref{inve3} and theorem \ref{char1} we obtain
\begin{eqnarray}
\nonumber
A^{-1} \otimes_{\mathbf{j}} B^{-1} &=& (\mathcal{A}_1^{-1} \e_1+ \mathcal{A}_2^{-1} \e_2) \otimes_{\mathbf{j}} (\mathcal{B}_1^{-1} \e_1+ \mathcal{B}_2^{-1} \e_2)\\
\label{second2}
&=&(\mathcal{A}_1^{-1} \otimes \mathcal{B}_1^{-1}) \e_1+(\mathcal{A}_2^{-1} \otimes \mathcal{B}_2^{-1})^{-1} \e_2.
\end{eqnarray}
Finally, since the right hand side of \eqref{first1} and \eqref{second2} are equal we get the thesis.
\end{proof}
\begin{remark}
One can reduce the tensor product of bicomplex matrices to the subset of hyperbolic matrices and define the product of $A\in \mathbb{D}^{n \times n}$ and $B\in \mathbb{D}^{m \times m}$ as in Definition~\ref{tensor}.
\end{remark}
 
 In fact, from theorem~\ref{char1}, we have:

\begin{lemma} If $A= \mathcal{A}_1 \e_1+ \mathcal{A}_2 \e_2 \in \mathbb{D}^{n \times n}$ and $B= \mathcal{B}_1 \e_1+ \mathcal{B}_2 \e_2 \in \mathbb{D}^{m \times m}$, where $\mathcal{A}_1, \mathcal{A}_2\in \mathbb R ^{n\times n}$ and $\mathcal{B}_1, \mathcal{B}_2\in\mathbb R^{m \times m},$ then the tensor product of $A$ and $B$ is:
$$A \otimes_{\mathbf{j}} B=(\mathcal{A}_1 \otimes \mathcal{B}_1) \e_1+ (\mathcal{A}_2 \otimes \mathcal{B}_2) \e_2,$$
with $ A \otimes B \in \mathbb{D}^{mn \times mn}$.
\end{lemma}
\begin{proof}
The proof follows from standard arguments.
\end{proof}
All other properties of the tensor product of bicomplex matrices then restrict to the set of hyperbolic matrices.

\subsection{Recovering Bicomplex Tensor components}
\label{recover}

We recall that in the complex case one can recover the components of a tensor product of a positive complex matrix from the outcome matrix, if the outcome matrix is state, as in~\cite{Alp-Lew}:

\begin{proposition}
\label{complex-recover}
If $M\in\C^{n\, m \times n \, m} $ is a positive complex matrix, and $M=M_{n}\otimes M_m$, then $M_n$ and $M_m$ can be recovered uniquely from $M$ using the following algorithm:
\begin{eqnarray}
\label{cplx-tens-comp}
d^* M_n c&= \sum_{k=1}^m (d \otimes f_k)^* M (c \otimes f_k), \, \forall c, d \in \mathbb C^ n,\\
\nonumber d^* M_m c &= \sum_{k=1}^n (e_k\otimes d)^* M (f_k\otimes c), \, \forall c, d\in \mathbb C^ m,
\end{eqnarray}

where $f_k$ and $e_k$ are basis of $\mathbb C^ m$ and $\mathbb C^ n$ respectively and the tensor product is the complex one.

\end{proposition}
We have a similar result in the bicomplex case and we write:
\begin{proposition} If $M\in\mathbb{BC}^{n\, m \times n \, m} $ is a hyperbolic positive matrix, and 
$M=A_n\otimes_j B_m,$ for some $A_n\in \mathbb{BC}^{n\times n}$ and $B_m \in \mathbb{BC}^{m\times m}$ then $A_n$ and $B_m$ can be recovered uniquely from $M$.
\end{proposition}
\begin{proof}
The proof is based on the idempotent representation and properties of the bicomplex tensor product. 
Indeed, this follows from the fact that we can write $M=\mathcal{M}_1 \e_1+\mathcal{M}_2 \e_2$ the problem reduces to decomposing the two complex matrices $\mathcal{M}_1\in \mathbb{C}^{n\, m \times n \, m}$ and $\mathcal{M}_2\in \mathbb{C}^{n\, m \times n \, m}$ in their respective tensorial components using Proposition~\ref{complex-recover}. 
From $\mathcal{M}_1= \mathcal{A}_1\otimes \mathcal{B}_1$ and $\mathcal{M}_2=\mathcal{A}_2\otimes \mathcal{B}_2$, then one obtains  $\mathcal{A}_1,\,\mathcal{A}_2\in \mathbb{C}^{n\times n}$ and
 $\mathcal{B}_1,\,\mathcal{B}_2\in \mathbb{C}^{m\times m}$ as in equation~\eqref{cplx-tens-comp}.
Then, using the formulae in the proof Theorem~\ref{char1}, the recovering of $A_n=\mathcal{A}_1 \e_1 + \mathcal{A}_2 \e_2$ and $B_m=\mathcal{B}_1 \e_1 + \mathcal{B}_2 \e_2$ follows.
\end{proof}

\subsection{Bicomplex Tensor Product and Digital Signal Processing}

Following~\cite{EAI, GCT, GCT2} one can write an algorithm of reduction of a special class of bicomplex DSP operations and pattern recognition. It is important to note the connection relating the structure of a given tensor matrix and the feasibility of efficient implementations of an algorithm involving it.
Following~\cite{GCT}, in the complex case, if we consider the operation on a set of $n$ inputs given by the matrix product $Y_n=C_nX_n$ when the matrix $C_n$ such that $C_n$ can be expressed as $C_n=A_s\otimes B_r$, a tensor product of smaller matrices, then the algorithm can be simplified as follows. 
From $C_n=A_s\otimes B_r=(A_sI_s)\otimes (I_rB_r),$ using the properties of the tensor product, one obtains the following rule 
$Y_n=C_nX_n=(P_{n,s}(I_r\otimes A_s)P_{n,r})(I_s\otimes B_r)X_n.$ 

These algorithms naturally generalize to the bicomplex case and the new bicomplex mathematical programming language will follow the same rules as above using bicomplex matrices. This application will be studied in a following paper.


\section{A Bicomplex Choi Theorem}
\label{bcplx-choi}
\setcounter{equation}{0}
In this section we describe the concept of completely hyperbolic positive bicomplex matrices and prove a bicomplex Choi Theorem using these notions.

\subsection{A Choi Theorem in the complex case}

First we recall the definition of positive maps in the complex case.
\begin{definition}
\label{map-cplx-pos}
	A linear map $ \phi: \mathbb{C}^{n \times n} \to \mathbb{C}^{m \times m}$ is hyperbolic positive if and only if for all positive matrices $A \in \mathbb{C}^{n \times n}$ we have $ \phi(A)$ is positive.
\end{definition}

One can extend a hyperbolic positive map $ \phi: \mathbb{C}^{n \times n} \to \mathbb{C}^{m \times m}$ to block matrices $A=(A_{jk}) \in( \mathbb{C}^{n \times n})^{N \times N}$, for each $ N \geq 0$, by
\begin{equation}
	\phi_N(A)=(\phi(A_{jk}))_{i,j=1}^N \in( \mathbb{C}^{m \times m})^{N \times N}
\end{equation}
\begin{definition}
	A map $ \phi$ is completely positive if for each $ N \geq 0$ all the maps $\phi_N$ are positive.
\end{definition}

We recall the following result, see \cite{A,C}:
\begin{theorem}[Choi's Theorem]
	\label{choi}
	Let $ \phi: \mathbb{C}^{n \times n} \to \mathbb{C}^{m \times m}$ be a completely positive map. Then we can write
	\begin{equation}
		\phi(A)= \sum_{i=1}^r V_i A (\overline{V_i})^t,
	\end{equation}
	for some $V_1,..., V_r \in \mathbb{C}^{m \times n}.$
\end{theorem}

We recall the following definition of finite quantum channels on a Hilbert space. 

\begin{definition} Let $\mathcal H$ be a finite dimensional Hilbert space and $\mathcal B(\mathcal H)$ be the space of operators that act on $\mathcal H$. We say that the trace preserving map $\mathcal E:\mathcal B(\mathcal H)\rightarrow \mathcal B(\mathcal H)$ is a quantum channel on the Hilbert space $\mathcal H$ iff there exists a set of operators $\{ E_a\,\}_a$ in $\mathcal B(\mathcal H)$ such that 
$$\mathcal E(\sigma)=\sum_a E_a \sigma E_a^{\dagger},$$
for any $\sigma\in\mathcal B(\mathcal H)$.
\end{definition}

In fact, Choi's Theorem states that completely positive maps in the complex case are finite quantum channels and we develop the same statements in the bicomplex case. 

\subsection{A Choi Theorem in the Bicomplex Case}
We now define a notion of positivity for linear bicomplex maps in order to obtain similar results. 
In the bicomplex case the Hilbert spaces above will become Hilbert modules and a description of bicomplex quantum channels follows.
\begin{definition}
\label{map-hyper-pos}
	A linear map $ \phi: \mathbb{BC}^{n \times n} \to \mathbb{BC}^{m \times m}$ is hyperbolic positive if and only if for all hyperbolic positive matrices $A \in \mathbb{BC}^{n \times n}$ we have $ \phi(A)$ is hyperbolic positive.
\end{definition}

One can extend a hyperbolic positive map $ \phi: \mathbb{BC}^{n \times n} \to \mathbb{BC}^{m \times m}$ to block matrices $A=(A_{jk}) \in( \mathbb{BC}^{n \times n})^{N \times N}$, for each $ N \geq 0$, by
\begin{equation}
	\phi_N(A)=(\phi(A_{jk}))_{j,k=1}^N \in( \mathbb{BC}^{m \times m})^{N \times N}
\end{equation}
\begin{definition}
	A map $ \phi$ is completely positive if for each $ N \geq 0$ all the maps $\phi_N$ are hyperbolic positive.
\end{definition}
In the bicomplex setting we use the complex Choi Theorem to give the following characterization of hyperbolic completely positive matrices, where we use the $^{*}$ bicomplex conjugate as in Definition~\ref{conj}. This theorem proves that, in fact, a bicomplex completely positive map is a finite bicomplex quantum channel.
\begin{theorem}
Let $ \phi: \mathbb{BC}^{n \times n} \to \mathbb{BC}^{m \times m}$ be a bicomplex completely positive map. We write $ \phi(A)= \phi_1(\mathcal{A}_1) \e_1+ \phi_2(\mathcal{A}_2) \e_2$, where $ \phi_\ell: \mathbb{C}^{n \times n} \to \mathbb{C}^{m \times m}$, $\ell=1,2$. Then we have
\begin{itemize}
\item[1)] The map $\phi$ is completely positive if and only if $ \phi_1$ and $ \phi_2$ are completely positive.
\item[2)] A completely positive map $\phi$ in this context can be written as
	\begin{equation}
	\label{ope}
	\phi(A)= \sum_{i=1}^r V_i A (V_i^{*})^t,
\end{equation}
for some $V_1,..., V_r \in \mathbb{BC}^{m \times n}$.
\end{itemize}

\end{theorem}

\begin{proof}
 To prove [1], let us consider $A_{jk}$ to be the block matrices of the matrix $A$. The map $ \phi$ is completely hyperbolic positive if $ \phi_N(A)=\phi(A_{jk})$ are hyperbolic positive for all $N \geq 0$. 
 By the idempotent decomposition we can write
	$$ \phi_N(A)= \phi_N(\mathcal{A}_1) \e_1+ \phi_N(\mathcal{A}_2) \e_2,$$
	with $ \phi_N(\mathcal{A}_i)= \phi(\mathcal{A}_{i,jk})$, where we denote by $ \mathcal{A}_{i,jk}$ the block matrices of the matrices $ \mathcal{A}_i$, for $i=1,2$. Since $\phi_N(A)$ is hyperbolic positive by Proposition \ref{equivalence} we get that $\phi_N(\mathcal{A}_i)$ are also hyperbolic positive. Therefore the maps $\phi_\ell$, with $ \ell=1,2$ are completely  hyperbolic positive.
	
	For the converse, if $\phi_N(\mathcal{A}_i)$ are hyperbolic positive  then by Proposition \ref{equivalence} we note that $\phi_N(A) $ is hyperbolic positive too. Then by Definition~\ref{map-hyper-pos} we get that $\phi(A)$ is completely hyperbolic positive.

From the first point of this theorem and \cite{C} we have
	\begin{eqnarray*}
		\phi(A)&=& \phi_1(\mathcal{A}_1) e_1+ \phi_2(\mathcal{A}_2)e_2\\
		&=&  \sum_{k=1}^r \mathcal{U}_k \mathcal{A}_1 (\overline{\mathcal{U}_k})^te_1+  \sum_{i=1}^r \mathcal{V}_k \mathcal{A}_2 (\overline{\mathcal{V}_k})^t e_2,
	\end{eqnarray*}
	where $ \mathcal{U}_i$, $ \mathcal{V}_i \in \mathbb{C}^{m \times n}$ with $1 \leq k \leq r$. We set
	$$ V_{k}:= \mathcal{U}_k e_1+ \mathcal{V}_k e_2.$$
	Therefore we have
	\begin{eqnarray*}
		\phi(A)&=& \sum_{i=1}^r \left(\mathcal{U}_k e_1+ \mathcal{V}_k e_2\right) \left(\mathcal{A}_1e_1+ \mathcal{A}_2e_2\right)[\left(\mathcal{U}_k e_1+ \mathcal{V}_k e_2\right)^*]^t\\
		&=& \sum_{k=1}^r V_k A (V_k^{*})^t.
	\end{eqnarray*}

\end{proof}

\begin{remark}
The form in formula \eqref{ope} is called Kraus decomposition and we observe that this representation is not unique.
\end{remark}

Using the idempotent representation for the tensor product and analogous results in the complex case the proof of the following statement readily follows.
\begin{proposition}
The bicomplex tensor product of two maps of the form \eqref{ope} has the same form.
\end{proposition}

In conclusion, this paper shows applications of the bicomplex Choi theorem to information theory and, as such, it is a fundamental first step of rewriting aspects of quantum mechanics from a bicomplex point of view.



\end{document}